\documentclass[12pt]{article}

\usepackage[margin=1in]{geometry}           
\usepackage{setspace}                       
\usepackage{times}                          
\usepackage{amsmath}                        
\usepackage{graphicx}                       
\usepackage{booktabs}                       
\usepackage{float}       
\usepackage{fancyhdr}                       
\usepackage[hidelinks]{hyperref}            

\usepackage[style=apa,
            backend=biber,
            natbib=true,
            uniquename=false,
            uniquelist=false]{biblatex}
\DeclareLanguageMapping{american}{american-apa}
\raggedright                                
\begin{document}

\title{Wait for It: A Component-Level Analysis of Latency Feedback Mechanisms for LLM Agents in VR}

\author{
Pratham Arora\\
Human Technology Interaction Lab \\
Plaksha University\\
Mohali, Punjab, India\\
\texttt{pratham.arora@plaksha.edu.in}
\and
Abhinav Lodha\\
Human Technology Interaction Lab \\
Plaksha University\\
Mohali, Punjab, India\\
\texttt{abhinav.lodha@plaksha.edu.in}
\and
Siddharth Siddharth\thanks{Corresponding author}\\
Human Technology Interaction Lab \\
Plaksha University\\
Mohali, Punjab, India\\
\texttt{siddharth.s@plaksha.edu.in}
}

\date{} 

\maketitle
\thispagestyle{fancy} 

\newpage
\begin{abstract}
Delays in AI responses pose a significant challenge to immersive Virtual Reality experiences. While previous research demonstrates that feedback can reduce the perception of waiting, we lack understanding of which types work best and why. We conducted a within-subjects study with 18 participants to systematically compare three isolated feedback approaches during VR agent interactions: verbal fillers (e.g., ``\textit{um...}," ``\textit{hmm...}"), visual indicators (loading spinners), and embodied gestures (thinking animations). Our results revealed that embodied gestures significantly outperformed visual loading indicators across multiple dimensions, making agents feel more lifelike and socially present. Interestingly, explicit visual feedback like loading spinners actually degraded perceived humanness and broke immersion by forcing users to shift from social to technical mental models. These findings suggest that natural, behavior-based feedback maintains the illusion of conversing with a thinking entity, while technical indicators expose the computational substrate. We present empirically-grounded design guidelines for creating more engaging AI-powered VR agents.
\end{abstract}


\newpage

\section{Introduction}
The convergence of Virtual Reality (VR) and Large Language Models (LLMs) has ushered in a new era of ``Intelligent Virtual Agents" (IVAs) capable of open-ended, free-form conversation. This big shift moves VR from an isolating experience toward a socially responsive environment where interaction mimics real-world dynamics~\parencite{elfleet2024investigating}. From educational tutors providing personalized feedback to virtual companions offering emotional support, LLM-powered avatars entails the creation of a seamless ``social presence" of being with another human being in a shared space.

However, one major flaw of LLM powered avatars threatens to break the illusion of presence: \textit{inference latency}. Modern LLMs, particularly those with sufficient reasoning capabilities to maintain a coherent persona and context (e.g., GPT-4, Gemini Pro), often require several seconds to process complex user input and generate a contextually appropriate response. In traditional text-based interfaces, such as messaging apps, users are accustomed to asynchronous communication; visual artifacts like ``typing bubbles" effectively bridge the temporal gap, signaling that a response is forthcoming. However, in face-to-face VR interactions, time perception is significantly compressed. The expectation of immediacy in spoken conversation is deeply ingrained in human psychology. A silence of 2--4 seconds—common in cloud-based LLM interactions—is perceived not merely as a delay, but as a ``break in presence." In human conversation, gaps longer than 200ms are often interpreted as hesitation, confusion, cognitive loading, or even social disengagement~\parencite{clark1996using}. When a VR avatar freezes or stares blankly for seconds at a time, the immersion collapses, and the user is starkly reminded that they are interacting with a software process rather than a social entity.

Prior research in Human-Computer Interaction (HCI) has proposed various strategies to mitigate this perceived latency. Many systems use ``fillers" like ``\textit{Hmm...}" or ``\textit{Let me see...}" to show they are thinking. While these work well for smart speakers, they can feel annoying and distracting in immersive settings like Virtual Reality. The experience can feel unnatural, if the filler does not match the emotional tone of the conversation. Alternatively, visual indicators (e.g., loading spinners, progress bars) borrow from 2D web and mobile design patterns to provide more transparency. While these explicitly inform the user of the system's operational status, they risk breaking the \textit{diegesis} of the virtual world—inserting a floating 2D User Interface (UI) element into a 3D narrative space, forcing the user to toggle between ``social" and ``technical" mental models.

Recent work by \textcite{elfleet2024investigating} demonstrated that a ``multimodal" approach combining verbal fillers, visual cues, and gestures significantly improved perceived response time compared to a no feedback of silence. While this establishes that \textit{some} feedback is indeed better than none, it obscures the specific cognitive contributions of each modality. Does the user feel better because the avatar used a filler, or because it shifted its weight? Did the visual spinner actually hurt immersion while the gesture helped it, or did they act synergistically? Understanding these component-level effects is crucial for designing interfaces that are not just quicker, but more socially credible.

To address this gap, we present a rigorous component-level analysis of latency feedback mechanisms in VR. We conducted a within-subjects study ($N=18$) where participants interacted with an LLM-powered avatar under three isolated feedback conditions: (1) \textbf{Verbal Fillers} (auditory only), (2) \textbf{Visual System Status} (explicit UI), and (3) \textbf{Embodied Gestures} (non-verbal behavioral cues), alongside a \textbf{no feedback condition} for control. 

Our results reveal a fascinating divergence between ``operational transparency" and ``social immersion". This work offers three primary contributions to the field of social Virtual Reality:
\begin{enumerate}
    \item \textbf{Empirical Decomposition:} We provide a component-level analysis isolating latency feedback modalities for LLM based interactions in VR. Contrasting with prior multimodal studies, we demonstrate that embodied non-verbal behavior is the primary driver of perceived animacy and immersion, independently outperforming verbal and visual strategies.
    \item \textbf{The Cost of Transparency:} We demonstrate that explicit system-status feedback can backfire in social VR contexts. Contrary to established usability principles that prioritize ``Visibility of System Status," we show that technical indicators (like loading spinners) break social immersion by reminding users they are interacting with a computational system. This suggests a fundamental trade-off in immersive agent design: transparency for trust versus coherence for presence.
    \item \textbf{Design Framework:} We provide a simple ``cheat sheet" for creators. Instead of using computer-like progress bars, we show how to use ``Embodied Wait States"—having the agent act like a human who is thinking—to make wait times feel shorter and more natural.
\end{enumerate}

\section{Related Work}

\subsection{Understanding Presence and User Experience in Immersive Environments}

When users interact with virtual characters, the quality of that experience depends fundamentally on how immersed they feel---a concept researchers call ``presence,'' defined as the psychological experience of ``being there'' in a virtual world. Presence is not purely a technical property; instead, it emerges from how well the virtual environment's features align with users' expectations and intentions~\parencite{triberti2025being}. In social VR contexts where a user interacts with an avatar, presence hinges on whether the avatar's behavior feels coherent and meaningful. Early foundational research by Slater and Steed introduced the concept of ``breaks in presence'' (BIPs)---moments when a user's perception suddenly shifts from the virtual world back to the physical world---and showed that more frequent BIPs correlate with lower presence ratings~\parencite{slater2000virtual}. \textcite{caldas2024breaking} recently demonstrated that when individual presence cues (such as embodiment, social presence, or narrative realism) are degraded, users' engagement and emotional responses decline measurably. This suggests that maintaining a cohesive virtual experience requires attention to every element that users perceive.

For virtual experiences, interface design plays a critical role. Research in diegetic design has identified two distinct UI approaches: diegetic elements, which are integrated into the virtual world's narrative and logic, versus non-diegetic elements, which exist outside the world and openly signal the underlying technical system~\parencite{saling2021diegetic,gottsacker2021diegetic}. \textcite{saling2021diegetic} compared these approaches in a VR game and found a clear preference for diegetic UIs, with trends toward higher presence ratings. \textcite{gottsacker2021diegetic} extended this finding to cross-reality interruptions, showing that fully and partially diegetic representations of social events yield higher place illusion and plausibility, lower interaction overhead, and fewer reported drops in presence compared to non-diegetic overlays or system notifications. This diegetic preference is not merely aesthetic---it reflects a fundamental principle: when interruptions or status information are rendered as in-world social events rather than as floating UI panels, users remain psychologically embedded in the virtual environment for longer. Conversely, a 2D loading spinner floating above an avatar functions as a non-diegetic HUD, which foregrounds the computational substrate and constitutes a designed break in presence. The diegetic design literature thus predicts that such cues can harm immersion even if they improve technical transparency, a prediction directly supported by the findings in this paper.

\subsection{Latency, Responsiveness, and Perceived Waiting in Human-Computer Interaction}

The passage of time in interactive systems is not merely a technical metric---it is a core element of user experience. Research in both desktop and mobile contexts has established that even small delays degrade satisfaction, with delays exceeding 2 seconds widely perceived as unacceptable in many interactive contexts. Notably, mobile users exhibit different tolerance thresholds: \textcite{arapakis2021impact} found that mobile web search users are approximately four times more tolerant of response latency than their desktop counterparts, although this tolerance collapses rapidly when delays exceed 7--10 seconds. This context dependency is important: mobile users may expect longer processing times due to network variability, yet they still form rapid judgments about system responsiveness.

One strategy for managing perceived delay is to provide active feedback during waiting periods. Research on visual feedback shows that dynamic progress indicators and animated loading bars can reduce perceived waiting time compared to static indicators or silence~\parencite{yu2023evaluation}. However, the benefits of such feedback are not universal. A recent study on the effect of visual feedback types on mobile app wait interfaces found that cartoon-style feedback led to shorter perceived waiting times than generic spinners, suggesting that the aesthetic and narrative quality of feedback matters~\parencite{chen2020effect}. The key insight is that feedback serves a communicative function: it reassures users that the system is working and gives them permission to wait.

In spoken dialogue and face-to-face interaction, the problem of latency becomes more acute. Clark established that in human conversation, gaps longer than 200 milliseconds are already interpreted as hesitation, confusion, or social disengagement~\parencite{clark1996using}. This sensitivity is deeply rooted in human turn-taking norms. Conversational turn transitions in spoken English occur at an average gap of roughly 239 milliseconds~\parencite{jacoby2024human}, a remarkably tight temporal window that reflects the collaborative nature of dialogue. When a speaker pauses to think or retrieve information, listeners do not sit in silence---instead, speakers deploy a variety of signaling behaviors to indicate that they are still holding the conversational floor and will soon respond. Backchannels (brief utterances like ``uh huh'' or head nods), fillers (non-lexical utterances like ``\textit{um}'' or ``\textit{let me see}''), and non-verbal behaviors (gaze shifts, hand movements) all serve this floor-holding function~\parencite{knudsen2020forgotten,vinciarelli2015words}. \textcite{knudsen2020forgotten} corpus analysis of Dutch and German conversation found that approximately 16--19\% of conversational turns consist of backchannels, and utterance-initial fillers are likewise ubiquitous. These minimal signals are easy to plan and produce, which is why they emerge so naturally and frequently in dialogue---they buy the speaker time for planning while maintaining engagement with the listener.

\subsection{Natural Feedback Mechanisms in Conversational Agents}

As conversational agents increasingly move into immersive and embodied contexts---such as virtual avatars in educational or social VR applications---the question of how to signal latency without breaking immersion becomes pressing. Prior research in Human-Robot Interaction (HRI) offers foundational insights. \textcite{wigdor2016improve} conducted a Wizard-of-Oz study with a physical robot (Robovie) and found that simple verbal fillers (e.g., ``\textit{Let me see}'') significantly improved the robot's perceived aliveness compared to silent delay, suggesting that auditory acknowledgment can mitigate the negative impact of latency. As conversational technology moved to screen-based agents, \textcite{boukaram2021mitigating} conducted a large-scale study ($N=360$) comparing verbal fillers against visual spinners in web-based dialogue agents. Their results revealed a subtle trade-off: contextualized fillers improved perceived speed but did not necessarily increase likeability relative to visual indicators.  Kum and Lee extended this work by contrasting verbal fillers with gestural fillers (e.g., chin stroking) in studies of digital human avatars, and found that gestural fillers were more effective at reducing perceived latency, suggesting that embodiment---the presence of an animated body---plays a crucial role in maintaining conversational floor~\parencite{kum2022effects}. \textcite{maslych2025mitigating} distinguished between ``Natural'' feedback (combinations of gestures and voice) and ``Artificial'' feedback (system-like signals), and found that the Natural condition mitigated long LLM delays (up to 6.5 seconds) more effectively, supporting the premise that socially meaningful behavior is more tolerant of delay than explicit system indicators.

Both verbal fillers and gestures are grounded in human communication norms. From the perspective of grounding theory (Clark \& Brennan), successful communication requires that conversational partners coordinate both content and process~\parencite{resnick1991perspectives,clark1996using}. This coordination is not achieved through explicit status messages; rather, it emerges through multimodal signals---voice, gaze, gesture, and timing---that constitute evidence of continued attention and engagement. \textcite{kendrick2023turn} recently demonstrated that in human face-to-face conversation, turn-taking coordination is fundamentally multimodal: gaze direction and manual gestures jointly regulate turn transitions, with averted gaze and unfinished gestures serving as powerful signals that the speaker intends to retain the floor. Building on this, \textcite{andrist2013conversational} showed that virtual agents exhibiting well-timed gaze aversion during conversational pauses are perceived as thinking, elicit more user disclosure, and are better able to hold the conversational floor compared to agents that stare continuously or avert gaze at inappropriate moments. These findings establish that silence combined with embodied thinking behavior---gaze aversion, micro-movements, posture shifts---already constitutes a recognizable floor-holding signal in human expectations. In this light, both verbal fillers and gestures can be understood as natural feedback: they leverage innate human communication patterns to signal continued cognitive engagement.

Visual system status indicators (such as loading spinners or progress bars) represent a fundamentally different category. These elements explicitly reveal the computational status of the underlying system---they are non-diegetic cues that place the agent's processing into the foreground of user attention. While such transparency is a cornerstone of usability heuristics, in immersive social environments this transparency comes at a cost.

\subsection{Embodied Conversational Agents: Perception, Anthropomorphism, and Responsiveness}

An embodied conversational agent (ECA) is a lifelike virtual human capable of producing speech, hand gestures, and facial expressions. The success of an ECA is not determined solely by dialogue quality or knowledge depth; instead, users' perception of the agent as a believable social entity---what researchers call anthropomorphism and social presence---heavily influences the interaction. Anthropomorphism refers to the human tendency to attribute human-like qualities (emotions, intentions, beliefs) to non-human entities~\parencite{damiano2018anthropomorphism}, while social presence refers to the sense of ``being with'' another agent and the degree to which that agent feels like a sentient being~\parencite{caldas2024breaking,wong2025review}.

Visual appearance is a significant driver of anthropomorphism. \textcite{thaler2020agent} compared autonomous ECAs to human-driven avatars (face-tracked representations) and found that increased visual humanness correlates with increased perceived humanness, though interestingly it also correlates with increased perceived eeriness (the uncanny valley effect). \textcite{nowak2003effect} foundational work showed that people respond socially to both human and computer-controlled virtual entities, and that the presence of a visual representation increases telepresence and social presence compared to no image at all. However, higher anthropomorphism does not automatically translate to better interaction. They also found a nuanced result: participants interacting with less-anthropomorphic avatars reported more copresence and social presence than those interacting with highly anthropomorphic images, suggesting that overly realistic avatars set up higher expectations that go unmet, leading to reduced presence when the agent's behavior does not match its appearance.

In conversational VR agents equipped with LLM capabilities, both familiarity (whether the agent resembles a known person) and knowledgeability (domain expertise) influence user perceptions of the agent as credible, trustworthy, and worth future engagement~\parencite{yang2025exploring}. These studies establish that users actively evaluate agents along multiple dimensions---not just on what they say, but on whether they respond appropriately, attend to the user's needs, and demonstrate coherent intentionality.

\section{Method}

\begin{figure*}[!htbp]
  \centering
  \includegraphics[width=\textwidth]{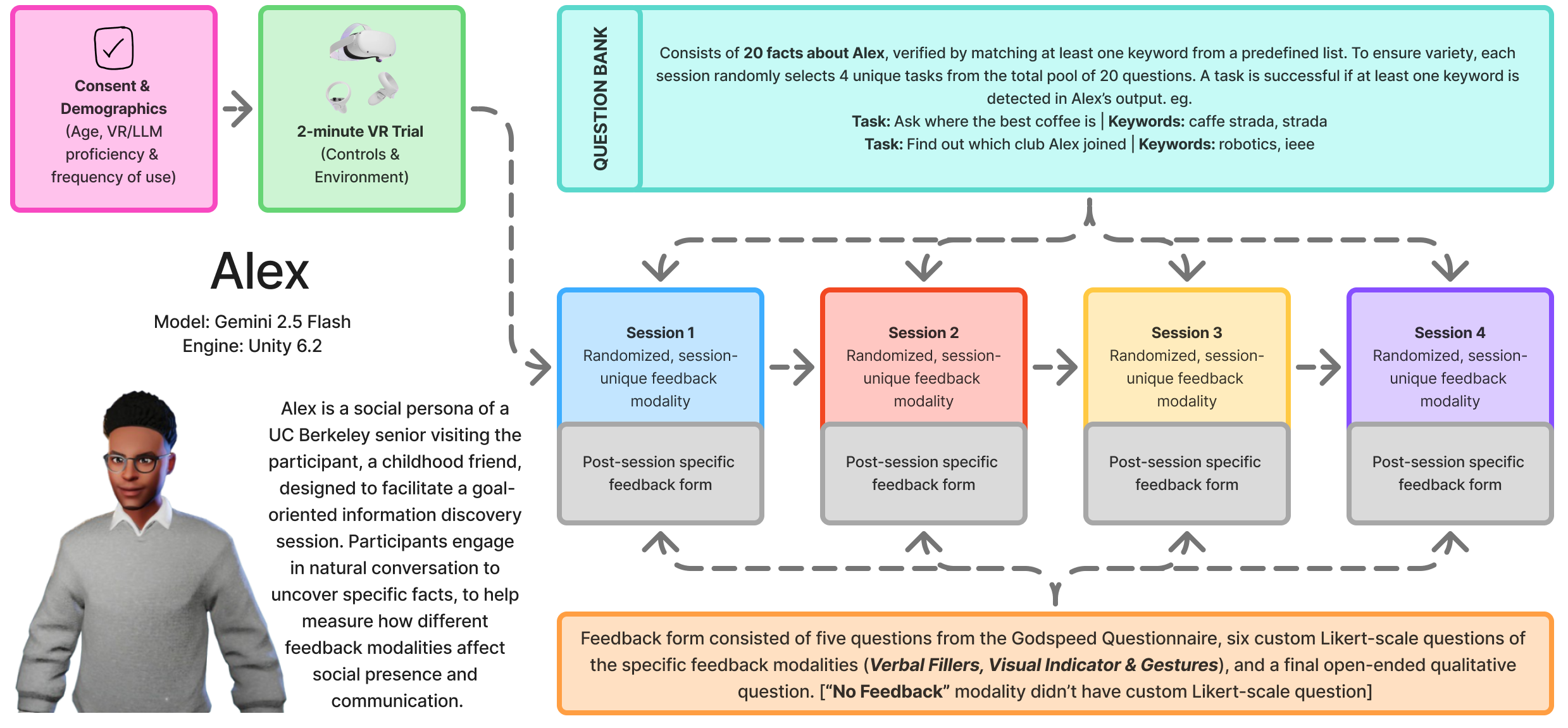}
  \caption{Experiment Overview}
  \label{fig:architecture}
\end{figure*}

\subsection{System Architecture}
The experimental environment was developed in \textbf{Unity 6 (Editor v6000.2.9f1)} and deployed on the \textbf{Meta Quest 2} headset via Oculus Air Link for high-fidelity PC-VR rendering (Figure~\ref{fig:architecture}). The virtual character, ``Alex" was created using a ReadyPlayerMe avatar. We gave him full-body movements (IK) and idle and speaking animations to make him look and feel more life-like. The setting was a casual cafe, chosen to make the interaction feel like a real-world encounter. In this scenario, Alex was participant's childhood friend who studies at UC Berkeley but has traveled to Mohali, India to visit the participant. The goal as the participant was to talk to Alex and learn a few specific facts about his life.

We utilized the \textbf{Google Gemini 2.5 Flash API} for intelligence. User speech was transcribed via Google Cloud STT, processed by the LLM (with regex cleaning to remove markdown/emojis), and synthesized via Google Cloud TTS. Average system round-trip latency was approximately 2.0 seconds ($SD=0.4s$). To enhance realism and enforce reliance on verbal cues, no visual transcriptions of the dialogue were displayed to the user.

\subsection{Participants}

We recruited 18 participants ($M_{age} = 21.28$ years, $SD = 2.42$, range: 19--30 years; 9 male, 9 female) following two pilot sessions conducted to refine the experimental procedure. The majority of participants (61.1\%, $n=11$) had no prior VR experience (first or second time using VR), while 27.8\% ($n=5$) had tried VR a handful of times, and 11.1\% ($n=2$) used VR occasionally (once every few months). All participants provided informed consent before beginning the study.

\subsection{Ethical Considerations}

This study was conducted in accordance with ethical research principles. All participants provided informed consent via digital waiver forms prior to participation. Participants were informed that their involvement would require wearing a VR headset and interacting with a 3D avatar in a virtual environment, engaging in conversations across multiple experimental conditions, completing questionnaires and surveys, and participating in a brief debriefing session. The total time commitment was approximately 30 minutes in a single session.

Data collection included automatically tracked data (response times and session logs with timestamps and conversation logs) and self-reported data (demographic information including age, gender, and VR experience level, as well as questionnaire responses regarding perceived latency, immersion, and user experience). 

To protect participant privacy, personal information was kept separate from research data. All data were stored securely on an encrypted, password-protected database with access limited to the research team. 

Participants were informed of potential risks including temporary symptoms such as motion sickness, nausea, dizziness, disorientation, eye strain, fatigue, headaches, and physical discomfort from wearing the VR headset. Participants were given the right to take breaks at any time or discontinue participation without penalty if they experienced discomfort. All participants confirmed their understanding that participation was voluntary, that they could withdraw at any time, that data would be kept confidential and used only for research purposes, and that they should inform the researcher immediately if they experienced any discomfort.

\subsection{Materials}

The study employed a Meta Quest 2 VR headset connected to a PC via Oculus Air Link. The virtual environment consisted of a cafe setting with a ReadyPlayerMe avatar (``Alex") capable of full-body inverse kinematics, idle animations, and speaking animations. The system utilized Google Gemini 2.5 Flash API for natural language processing, Google Cloud Speech-to-Text for transcription, and Google Cloud Text-to-Speech for audio synthesis.

Four feedback conditions were implemented:
\begin{enumerate}
    \item \textbf{No Feedback:} The agent remained in a standard idle state with no audio or visual indication during processing.
    \item \textbf{Visual Indicator:} A 2D loading spinner appeared above the agent's head 0.3s after the user finished speaking.
    \item \textbf{Verbal Fillers:} The agent uttered a randomized, context-agnostic non-lexical filler (e.g., ``\textit{Hmm}," ``\textit{Let me see}," ``\textit{Uh}") selected at random from a predefined set of 24 pre-recorded utterances (see Supplementary Materials, Appendix C for the complete inventory) approximately 0.5s after the user's speech ended. These fillers were not dynamically generated based on conversational content or emotional tone, representing a methodological choice to isolate the effect of verbal acknowledgment without introducing additional processing latency that we discuss in Section~5 (Limitations).
    \item \textbf{Gestures:} The agent performed one of two ``thinking" animations—tapping fingers on the table or folding his arms and looking around.
\end{enumerate}

\subsection{Procedure}

\begin{figure*}[!htbp]
  \centering
  \begin{minipage}[b]{0.45\textwidth}
    \centering
    \includegraphics[width=\linewidth]{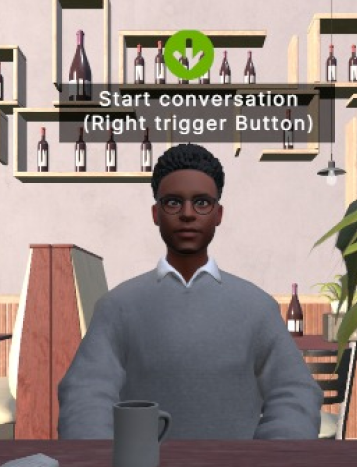}\\
    (A)
  \end{minipage}
  \hfill
  \begin{minipage}[b]{0.45\textwidth}
    \centering
    \includegraphics[width=\linewidth]{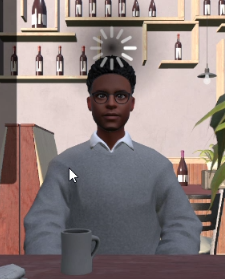}\\
    (B)
  \end{minipage}
  
  \vspace{0.5cm}
  
  \begin{minipage}[b]{0.45\textwidth}
    \centering
    \includegraphics[width=\linewidth]{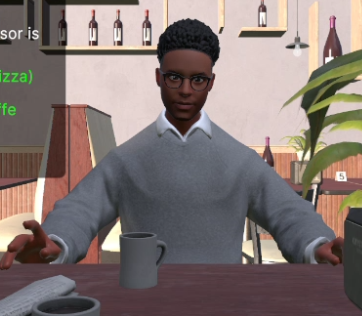}\\
    (C)
  \end{minipage}
  \hfill
  \begin{minipage}[b]{0.45\textwidth}
    \centering
    \includegraphics[width=\linewidth]{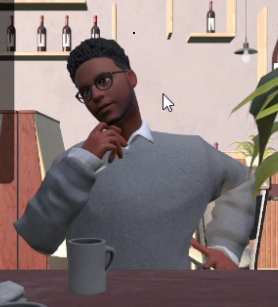}\\
    (D)
  \end{minipage}
  \caption{Virtual environment and feedback mechanisms used in the study. (A) Cafe setting where participants conducted conversations with the virtual agent Alex. (B) Visual feedback condition displaying a loading spinner above the agent's head. (C) Gestural feedback: finger-tapping animation on the table surface. (D) Gestural feedback: contemplative pose with folded arms. Note: The verbal filler condition is not shown as it involved auditory feedback only (randomized non-lexical utterances such as ``\textit{Hmm},'' ``\textit{Let me see},'' or ``\textit{Uh}'') with no visual changes to the agent's appearance.}
  \label{fig:environment_feedback}
\end{figure*}

We employed a within-subjects design to evaluate the impact of different feedback modalities on user experience during VR-based human-agent interactions. The experimental flow was designed to minimize learning effects while ensuring participants were fully accustomed to the VR environment before data collection began. The process followed a linear progression through three main phases:

\begin{enumerate}
    \item \textbf{Onboarding:} Participants first provided informed consent and completed a demographic survey. To mitigate the ``novelty effect" of VR, they underwent a two-minute trial session. During this time, they practiced using the VR controllers and navigated the environment to become comfortable with the spatial layout and the agent's presence (Figure~\ref{fig:environment_feedback}A).
    \item \textbf{Experimental Trials:} Participants completed four interaction sessions (no feedback, Verbal, Visual, Gesture) in a randomized order. Each session lasted until the participant successfully extracted 4 unique facts (approx. 5--8 minutes).
    \item \textbf{Data Collection:} In each session, participants were tasked with extracting four unique ``Life Facts" from the agent. These facts were randomly assigned and varied between sessions to ensure no overlap. Immediately following each session, participants completed a post-session survey, which included five questions chosen from Godspeed Questionnaire, six questions from Likert scale based custom questionnaire about each feedback modality and finally a open ended feedback question. 
\end{enumerate}

\section{Results}

We analyzed subjective perceptions of the agent across four feedback conditions using the Godspeed Questionnaire and custom measures. Given the relatively small sample and ordinal data, we employed non-parametric tests (Friedman tests for global effects, Wilcoxon signed-rank tests for pairwise comparisons) alongside standard paired-sample $t$-tests for transparency. Pairwise $p$-values were Holm-corrected within each dimension ($p_{adj}$).

\subsection{Latency Consistency Verification}

To verify that network-dependent LLM latency fluctuations did not systematically bias our feedback condition comparisons, we analyzed the actual response times logged for each condition. The mean response latencies were: No Feedback ($M = 1.94s$, $SD = 0.58s$), Gestures ($M = 1.98s$, $SD = 0.62s$), Visual ($M = 2.02s$, $SD = 0.51s$), and Verbal ($M = 1.92s$, $SD = 0.60s$). A Friedman test showed no significant difference in actual latency across conditions ($\chi^2(3) = 1.13$, $p = 0.769$), confirming that our randomized session ordering successfully distributed network variance evenly across all feedback types. This allows us to attribute perceptual differences to the feedback mechanisms themselves rather than to systematic latency variations.

\subsection{Agent Animacy and Anthropomorphism}

A Friedman test revealed significant main effects for both animacy ($\chi^2(3) = 9.44$, $p = 0.024$) and anthropomorphism ($\chi^2(3) = 9.66$, $p = 0.022$). These dimensions capture whether users perceived the agent as alive and as having human-like qualities---core aspects of social presence.

\subsubsection{Animacy}

Pairwise comparisons showed that embodied gestures ($M = 4.22$, $SD = 0.65$) significantly enhanced perceived animacy relative to both the visual spinner ($M = 3.39$, $SD = 1.04$; $t(17) = 3.07$, $p = 0.007$, $p_{adj} = 0.034$) and verbal fillers ($M = 3.44$, $SD = 0.98$; $t(17) = 3.76$, $p = 0.002$, $p_{adj} = 0.009$). The comparison against no feedback ($M = 3.67$, $SD = 0.97$) trended toward significance ($p = 0.056$). Table~\ref{tab:pairwise_comparisons} provides a comprehensive summary of all statistical comparisons across metrics.

Users' quantitative preference for gestures reflected a clear experiential difference. One participant captured the essence: \textit{``The gestures felt very natural and greatly enhanced my experience.''} Another simply stated, \textit{``The hand gestures made him more human-like.''} These remarks underscore the quantitative finding: when delay is filled by a recognizable thinking gesture, users interpret it as the agent actively considering the question rather than the system being unresponsive or broken.

\begin{figure*}[!htbp]
  \centering
  \includegraphics[width=1.0\textwidth]{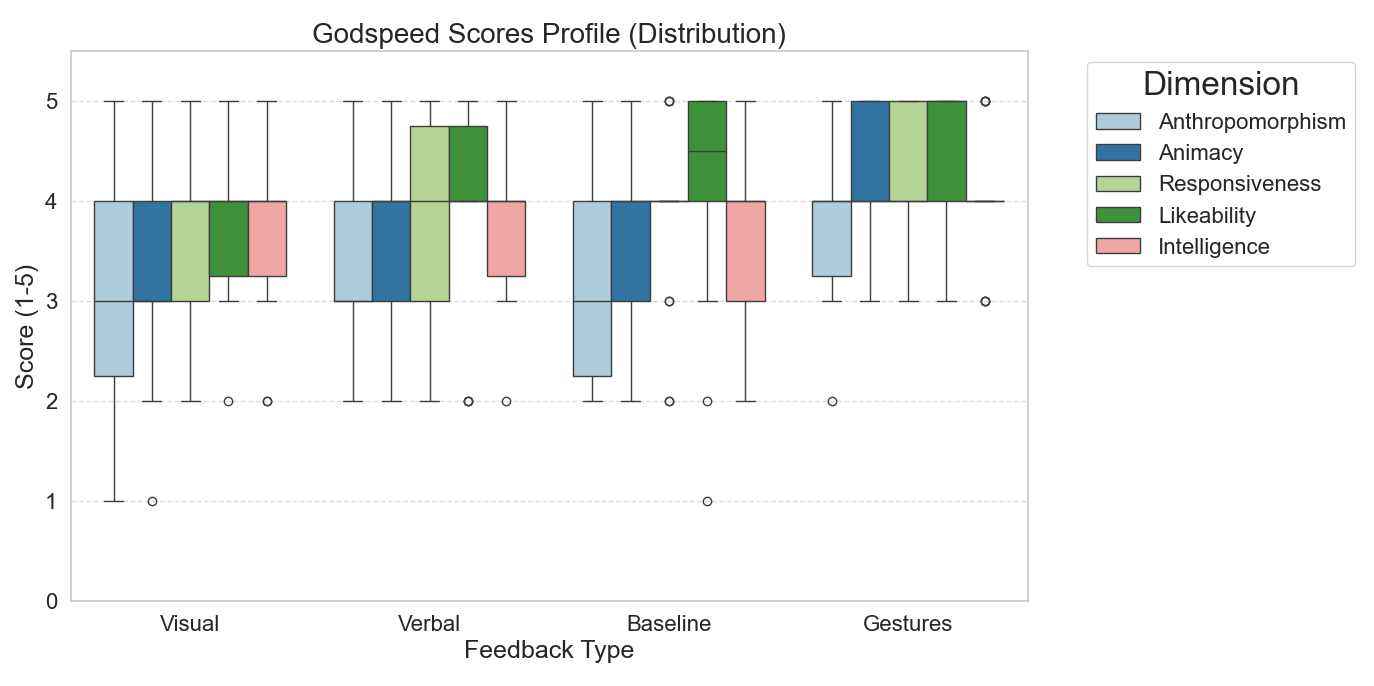} 
  \caption{Godspeed scores profile showing mean ratings across all dimensions. Gestures condition consistently scores highest on social dimensions (Animacy, Anthropomorphism, Likeability). Scale: 1 = Negative anchor (e.g., Machinelike, Stagnant, Unfriendly, Unintelligent, Apathetic); 5 = Positive anchor (e.g., Humanlike, Lively, Friendly, Intelligent, Responsive).}
  \label{fig:godspeed_results}
\end{figure*}

\subsubsection{Anthropomorphism}

The visual spinner condition scored significantly lower ($M = 3.00$, $SD = 1.08$) than gestures ($M = 3.83$, $SD = 0.79$; $t(17) = 3.59$, $p = 0.002$, $p_{adj} = 0.014$). This is a striking finding: by explicitly revealing the computational process at work, the visual cue inadvertently stripped the agent of human-like qualities.

Qualitative feedback illuminated this paradox. When shown the spinner, participants felt it contradicted the social premise of the interaction: \textit{``Loading icon didn't feel authentic---took away authenticity from the environment.''} Another noted that the spinner \textit{``can be replaced with something more humanlike.''} In contrast, gestures preserved the illusion of an entity who was thinking, not a system that was computing.

\begin{table*}[ht!]
\caption{Summary of Top Statistical Comparisons (paired $t$-tests). $p_{adj}$: Holm-corrected within each dimension; asterisks (*) denote $p_{adj} < 0.05$. Bold values indicate the higher-rated condition in each comparison. See Supplementary Materials (Appendix A) for complete pairwise results.}
\label{tab:pairwise_comparisons}
\centering
\small
\begin{tabular}{@{}lp{4.5cm}cccccc@{}}
\toprule
\textbf{Metric} & \textbf{Comparison} & \textbf{Group 1} & \textbf{Group 2} & \textbf{\textit{t}(17)} & \textbf{\textit{p}} & \textbf{\textit{p}$_{adj}$} \\
 & & \textbf{M (SD)} & \textbf{M (SD)} & & & \\
\midrule
Animacy & Gestures vs Verbal & \textbf{4.22 (0.65)} & 3.44 (0.98) & 3.76 & 0.002 & \textbf{0.009*} \\
Anthropomorphism & Gestures vs Visual & \textbf{3.83 (0.79)} & 3.00 (1.08) & 3.59 & 0.002 & \textbf{0.014*} \\
Animacy & Gestures vs Visual & \textbf{4.22 (0.65)} & 3.39 (1.04) & 3.07 & 0.007 & \textbf{0.034*} \\
Intelligence & No Feedback vs Gestures & 3.61 (0.92) & \textbf{4.06 (0.64)} & $-$2.68 & 0.016 & 0.096 \\
Likeability & Gestures vs Visual & \textbf{4.44 (0.70)} & 3.89 (0.83) & 2.26 & 0.037 & 0.222 \\
Animacy & No Feedback vs Gestures & 3.67 (0.97) & \textbf{4.22 (0.65)} & $-$2.05 & 0.056 & 0.224 \\
Anthropomorphism & No Feedback vs Gestures & 3.33 (1.14) & \textbf{3.83 (0.79)} & $-$2.03 & 0.058 & 0.290 \\
Likeability & Gestures vs Verbal & \textbf{4.44 (0.70)} & 3.94 (1.00) & 1.84 & 0.083 & 0.413 \\
Anthropomorphism & Gestures vs Verbal & \textbf{3.83 (0.79)} & 3.39 (0.92) & 1.81 & 0.088 & 0.352 \\
Intelligence & Gestures vs Visual & \textbf{4.06 (0.64)} & 3.78 (0.88) & 1.76 & 0.096 & 0.481 \\
\bottomrule
\end{tabular}
\end{table*}

\subsection{Likeability and Perceived Intelligence}

Pairwise analysis showed a preference for gestures ($M = 4.44$, $SD = 0.70$) over the visual spinner ($M = 3.89$, $SD = 0.83$; $t(17) = 2.26$, $p = 0.037$), suggesting an affective preference for social congruence over technical clarity; however, this difference did not survive correction ($p_{adj} = 0.222$), and the Friedman test for likeability was not significant ($\chi^2(3) = 4.75$, $p = 0.191$). Intelligence ratings showed a similar trend favoring gestures ($M = 4.06$, $SD = 0.64$) over no feedback ($M = 3.61$, $SD = 0.92$; $t(17) = -2.68$, $p = 0.016$), though this was not significant after correction ($p_{adj} = 0.096$; Friedman $\chi^2(3) = 5.72$, $p = 0.126$). Intelligence perception thus proved less sensitive to latency-masking strategies than social presence measures.

\subsection{Distraction and Perceived Naturalness}

Participants rated how distracting the waiting period felt (1 = Not distracting, 5 = Very distracting). A Wilcoxon signed-rank test showed that embodied gestures ($M = 1.89$, $Mdn = 2.0$) were significantly less distracting than visual spinners ($M = 2.67$, $Mdn = 3.0$; $W = 4.0$, $p = 0.008$) and verbal fillers ($M = 2.78$, $Mdn = 3.0$; $W = 10.5$, $p = 0.012$). 

On the complementary measure of perceived naturalness (how natural the waiting felt), gestures again prevailed, aligning with the finding that natural feedback mechanisms feel less intrusive. Verbal fillers fell between gestures and the spinner, suggesting they occupied a middle ground—socially intentional enough to not feel purely mechanical, yet prone to misalignment with conversational tone.

Participants explicitly noted this tension: \textit{``Filler words were fine, but conversation seemed little jittery.''} Another observed, \textit{``The fillers could be made more natural.''} A third reflected, \textit{``The filler words, wouldn't be the best.''} These remarks highlight a critical limitation: context-agnostic fillers, while grounded in human communication, risk tonal discord when they clash with the emotional or semantic tenor of the ongoing conversation.

\subsection{Perception of Latency Masking}

We asked participants to rate agreement with: \textit{``The feedback made the waiting time feel shorter''} (1 = Strongly Disagree, 5 = Strongly Agree). Embodied gestures received the highest median rating ($Mdn = 4.0$), indicating that users felt the wait was effectively compressed when the agent remained active. Verbal fillers achieved a median of 3.0---suggesting moderate effectiveness at masking delay. Visual spinners were rated lowest ($Mdn = 2.0$; gestures vs. spinner: Wilcoxon $W = 26.0$, $p = 0.026$), suggesting that explicit system-status indicators often \textit{highlighted} the delay rather than disguising it, making users more conscious of the waiting period.

This pattern supports a key insight: naturalistically grounded feedback (gestures) serves a temporal masking function more effectively than explicit system-status indicators (spinners) or decontextualized social signals (verbal fillers).

\subsection{Influence of Prior VR Experience}

Given that the majority of our participants (61.1\%, $n=11$) had minimal or no prior VR experience (first or second time using VR), while 27.8\% ($n=5$) had tried VR a handful of times and 11.1\% ($n=2$) used VR occasionally, we examined whether VR familiarity moderated the effectiveness of different feedback mechanisms. We compared Godspeed ratings between VR novices and those with at least some prior exposure.

While VR novices showed slightly lower baseline ratings for agent likeability ($M = 3.95$, $SD = 0.71$) compared to experienced users ($M = 4.36$, $SD = 0.52$; per-participant means), this difference did not reach significance ($t(16) = -1.30$, $p = 0.21$). Critically, the \textit{relative ordering} of feedback effectiveness remained consistent across experience levels: embodied gestures consistently outperformed visual indicators in both groups (Novices: $\Delta M = 0.45$; Experienced: $\Delta M = 0.71$). This suggests that the preference for diegetic, socially-grounded feedback over explicit system indicators is not merely a novelty effect, but reflects a fundamental human expectation for natural communication during face-to-face (even virtual) interaction.

\section{Discussion}

\begin{figure*}[!htbp]
  \centering
  \includegraphics[width=0.95\textwidth]{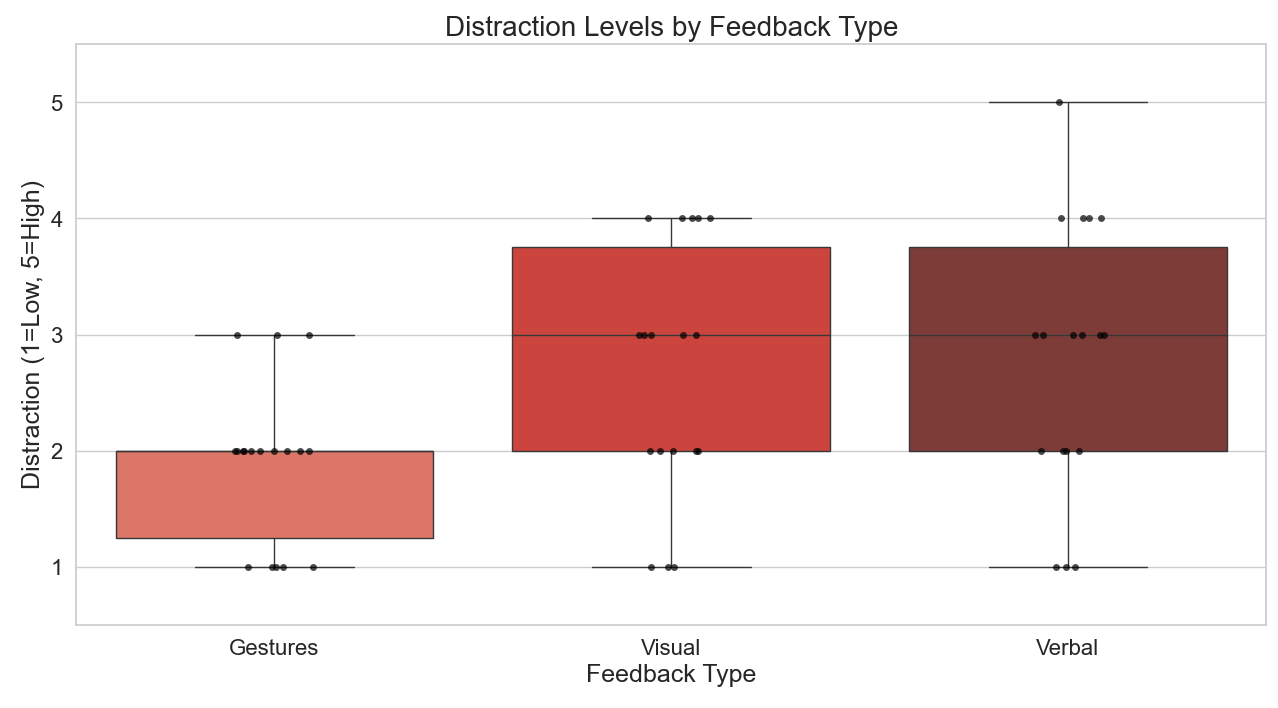}\\
  \vspace{0.3cm}
  (A)
  
  \vspace{0.8cm}
  
  \includegraphics[width=0.95\textwidth]{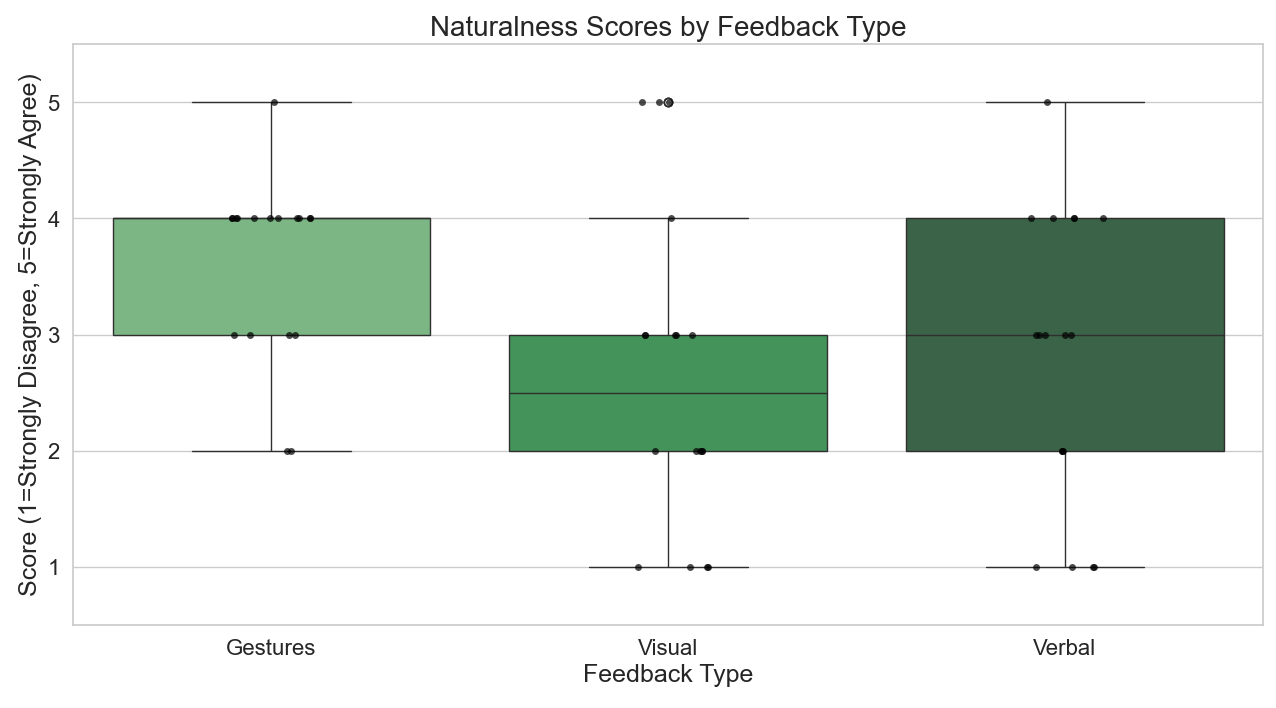}\\
  \vspace{0.3cm}
  (B)
  \caption{Subjective Experience Ratings. (A) Distraction Levels (Lower is better): Gestures are significantly less distracting than visual or verbal feedback. (B) Naturalness Scores (Higher is better): Gestures create the most natural waiting experience.}
  \label{fig:distraction}
\end{figure*}

Our findings challenge a fundamental assumption in HCI design: that transparency always improves user experience. When developers show users ``what the system is doing,'' they often assume this helps. In VR, we found the opposite. While prior work advocates for transparency (letting users know the computer is processing), our results show that in immersive social environments, transparency can paradoxically \textit{degrade} the user experience. This tension drives three main insights.

\begin{figure*}[!htbp]
  \centering
  \includegraphics[width=1.0\textwidth]{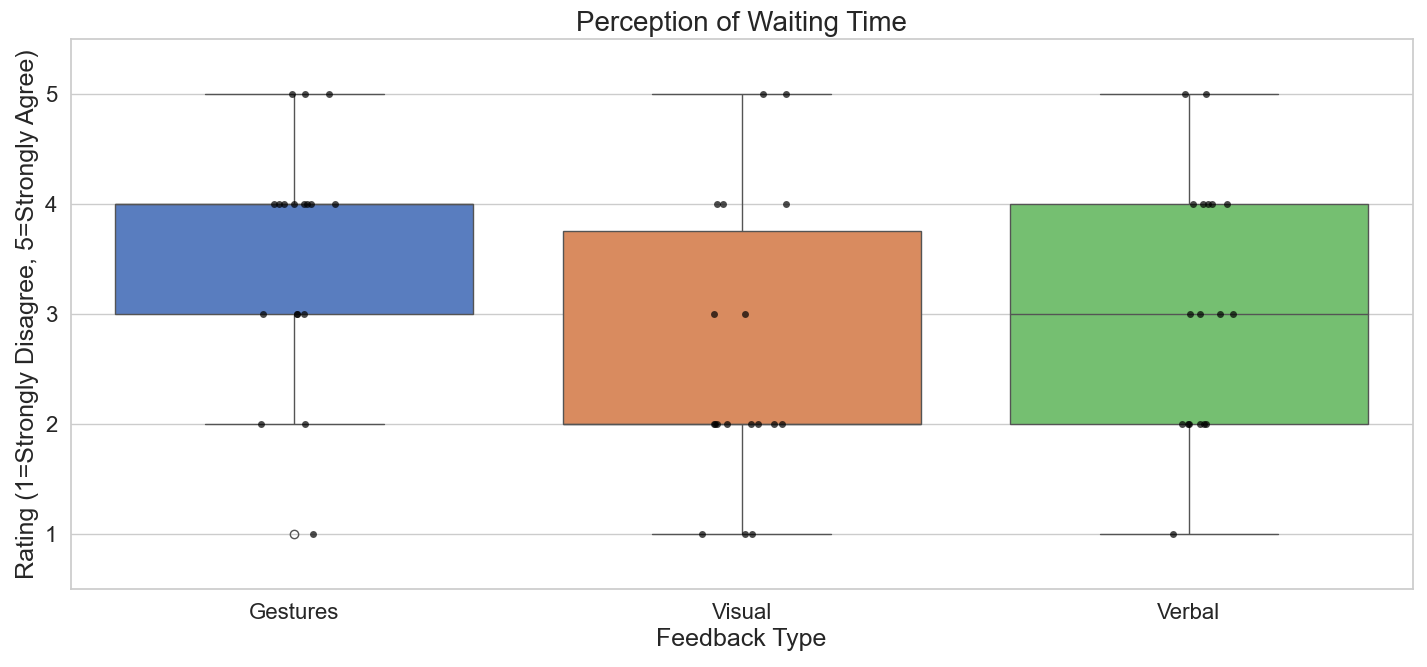}\\
  \vspace{0.3cm}
  (A)
  
  \vspace{0.8cm}
  
  \includegraphics[width=1.0\textwidth]{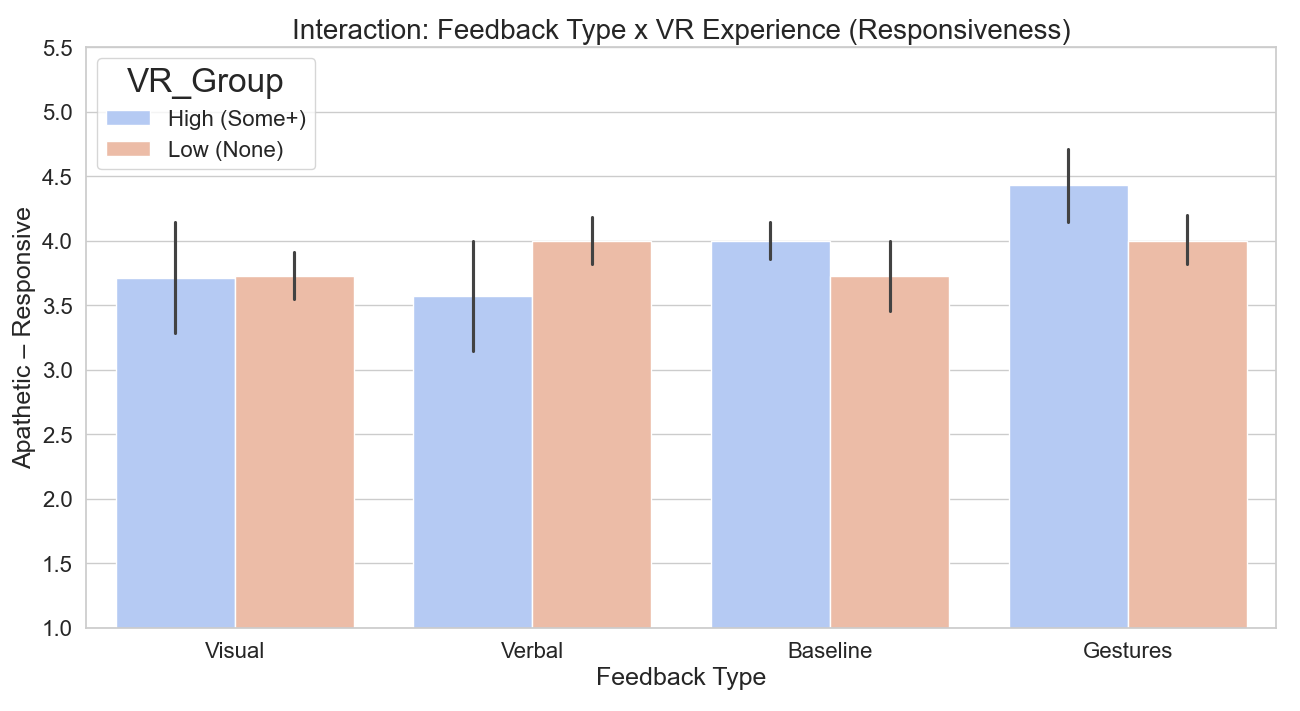}\\
  \vspace{0.3cm}
  (B)
  \caption{Temporal Perception and Interaction Effects. (A) Perception of Waiting Time: Gestures make the wait feel shorter compared to Visual or Verbal feedback. (B) Interaction Effect: VR experience modulates feedback effectiveness on perceived responsiveness.}
  \label{fig:temporal}
\end{figure*}

\subsection{The Diegetic Gap: Why Transparency Breaks Immersion in VR}

The visual spinner condition showed a statistically significant drop in perceived humanness (Anthropomorphism, $p = 0.002$) and lifelikeness (Animacy, $p = 0.007$). This finding directly contradicts a cornerstone of interaction design: Nielsen's heuristic of ``Visibility of System Status,'' which recommends that systems always keep users informed about what is happening.

In a traditional interface---a web page, a mobile app, a command-line tool---a loading spinner is genuinely helpful. It reduces uncertainty. Users want to know: ``Is the system working or did I break something?''

In VR, the same design pattern produces a radically different effect. When a spinner hovers above an avatar's head in a 3D caf\'{e}, it violates what film theorists call the ``diegesis''---the story world. A 2D floating icon tears users out of the social narrative and forces them to switch mental modes: from ``I'm talking to a person'' to ``I'm monitoring a process.'' This toggle is so jarring that users report the agent feels less human afterward, even though they now have better information about system status.

Our data suggests that VR users prioritize \textit{diegetic consistency} (staying ``in the story'') over explicit system status visibility. This echoes \textcite{maslych2025mitigating} distinction between ``natural'' (socially meaningful) feedback and ``artificial'' (system-level) feedback, but extends it with a clear cost-benefit trade-off: gaining technical clarity comes at the direct expense of social presence. In immersive environments, the user's sense of ``Is this a person or a machine?'' matters more than ``Is the system working?''

\subsection{Gestures as Social Keep-Alive Signals}

Embodied gestures succeed because latency in human-AI interaction is fundamentally a \textit{social event}, not a technical delay. Humans don't freeze while thinking---we shift weight, avert gaze, tap fingers. These conversational ``floor-holding'' signals indicate engagement and request patience. By replicating these cues, avatars reframe ``the LLM is computing'' into ``the agent is thinking,'' transforming latency from system failure into character expression.

This aligns with grounding theory: mutual understanding emerges through non-verbal signals (backchannels, gaze, gestures) rather than explicit status messages~\parencite{resnick1991perspectives,andrist2013conversational}. Participants recognized this: ``I didn't mind waiting when he looked like he was thinking; it felt polite to wait.'' Embodied behavior doesn't distract from latency---it reinterprets it as natural social interaction.

\subsection{Implications for AR/VR and Conversational AI}

These findings extend to any conversational interface. For mobile AR and partially-virtual agents, avoid explicit system UI (spinners, progress bars) for delays under 3--4 seconds. Use behavioral cues (head tilts, gaze shifts, gestures) that maintain the social frame. For voice assistants, verbal fillers must be contextually aware, reflecting the user's emotional tone rather than using randomized utterances.

The core principle: \textbf{behavioral congruence trumps technical transparency}. Socially appropriate responses (gesture, gaze, tone) maintain engagement despite latency.

\subsection{Design Framework: Embodied Wait States}

For developers building social VR, AR, and mobile experiences, we propose the \textbf{Embodied Wait State} framework as an alternative to traditional ``loading states'':

\begin{enumerate}

\item \textbf{Hide the System, Reveal the Character.} For conversational delays under 3--4 seconds, avoid 2D UI overlays (spinners, progress bars). The cost to perceived humanness outweighs the benefit of technical transparency. Instead, use behavioral cues.

\item \textbf{Embody the Wait.} Implement procedural ``thinking behaviors'': shifting weight, gazing upward thoughtfully, tapping fingers, or adjusting posture. These should trigger immediately upon detecting user silence and continue throughout the processing period.

\item \textbf{Silence + Gesture $>$ Random Fillers.} If verbal fillers cannot be emotionally contextual, omit them. A silent, thoughtful pause with appropriate gesture is more credible than a misaligned vocal filler.

\item \textbf{Test for Diegetic Consistency.} Before deploying any feedback mechanism, ask: ``Does this feel like part of the character's world, or like a system overlay?'' If it feels like the latter, redesign it.

\end{enumerate}

This framework is immediately implementable. Most VR/AR engines (Unity, Unreal) already support procedural animation and state machines; the design shift is conceptual, not technical.

\section{Limitations and Future Work}

Our study faced several constraints that frame our findings and point toward future research directions:

\begin{itemize}
    \item \textbf{Network-Dependent Latency Variance:} We relied on a cloud-based LLM (Gemini 2.5 Flash) via API, which introduced response time fluctuations ranging from 1.5s to 4.0s due to network conditions. While our session randomization successfully distributed this variance evenly across conditions (Friedman test: $\chi^2(3) = 1.13$, $p = 0.769$), preventing systematic bias, this variability may have weakened the precision of causal claims about feedback effectiveness. Future work should use on-device models (e.g., edge-deployed LLMs) to achieve millisecond-level latency control, enabling more precise testing of perceptual thresholds.
    
    \item \textbf{Context-Agnostic Verbal Fillers:} To isolate the effect of verbal acknowledgment without introducing confounding processing delays, we implemented verbal fillers as randomized utterances from a fixed list (e.g., ``\textit{Hmm}," ``\textit{Let me see}," ``\textit{Uh}"). This meant fillers could not adapt to the emotional tone or semantic content of the user's prior utterance. Due to a logging error, the specific filler played on each turn was not recorded, so effects of individual fillers could not be analyzed. Participants explicitly noted this limitation, reporting that fillers sometimes felt ``jittery" or tonally mismatched. Future implementations should explore lightweight, low-latency sentiment classifiers or edge-optimized language models to generate contextually appropriate fillers in real-time (e.g., empathetic acknowledgments for serious topics, enthusiastic responses for casual exchanges).
    
    \item \textbf{Sample Homogeneity and VR Novelty:} Our participant sample consisted predominantly of university students ($M_{age} = 21.3$ years), with 61.1\% reporting minimal or no prior VR experience. While we found that feedback preferences were consistent across VR experience levels (gestures outperformed visual cues in both novice and experienced groups), the lack of age and cultural diversity limits generalizability. Older adults or cross-cultural populations might exhibit different tolerance thresholds for silence, or different interpretations of gestural cues. Additionally, the novelty effect of VR for our sample may have elevated baseline engagement, potentially masking subtler differences between conditions.
    
    \item \textbf{Animation Fidelity:} The gestural animations were procedurally generated using standard inverse kinematics. Higher-fidelity motion capture data or real-time eye-tracking (allowing the agent to react to the user's gaze direction) might yield even stronger effects for embodied feedback conditions.
\end{itemize}

\section{Conclusion}

In this paper, we decomposed the black box of multimodal latency feedback to evaluate its individual components' impact on user immersion in VR. Explicit visual indicators, while functional, distract users and significantly reduce the sense that the agent is human-like. In contrast, subtle, embodied gestures successfully mask latency while enhancing the agent's perceived animacy.

As VR agents become ubiquitous in our lives, handling their latency and processing time will be just as critical as generating the words themselves. Our results advocate for a paradigm shift from technical wait states to \textbf{Embodied Wait States}—where the agent remains socially present, even when cognitively absent. By treating latency as a moment for character performance rather than a system loading error, designers can preserve the magic of immersion.

\section*{Acknowledgments}

We thank the participants for their time and valuable feedback. We also extend our gratitude to our research mentors for their guidance throughout this project, and to those who facilitated the VR hardware logistics.

\textbf{LLM Usage Statement.} We declare that large language models (LLMs) were used exclusively for language editing and stylistic improvements in this manuscript. They did not contribute to the conceptual, methodological, or experimental aspects of the work.

\textbf{Ethics Statement.} This work adheres to ethical research practices. The study was approved by our institution's ethics committee prior to data collection. Demographic and conversational data were collected from 18 participants. Prior to participation, all individuals were informed of the study's purpose and procedure. They provided digital informed consent before accessing the tasks and were explicitly notified of their right to withdraw at any time. No sensitive personal data was collected. All response data are anonymized and will be destroyed three months after the paper is accepted.

\printbibliography

\end{document}